\documentclass{iopart}
\usepackage[dvips]{graphicx}
\usepackage[english]{babel}

\renewcommand{\vec}{\bi} %vektor als fettdruck
\begin{document}

\title{Two-band ferromagnetic Kondo-lattice model for local-moment half-metals}
\author{M. Kreissl and W. Nolting}
\address{Humboldt University Berlin, Newtonstrasse 15, 12489 Berlin, Germany}
\eads{kreissl@nld.ds.mpg.de}

\begin{abstract}
We introduce a two-band Kondo-lattice model to describe ferromagnetic half-metals with local magnetic moments. In a model study, the electronic and magnetic properties are presented by temperature dependent magnetization curves, band-structures, spin polarizations and plasma  frequencies. These are obtained from numerically evaluated equations, based on the single-electron Green functions. We show that the mutual influence between the itinerant electrons and the local magnetic moments is responsible for several phase transitions of the half-metals, namely  first and second order magnetic phase transitions, as well as half-metal to semiconductor and half-metal to semimetal transitions.
\end{abstract}

\submitto{\JPCM}
%\noindent{\it Keywords\/}: half-metal, semimetal, semiconductor, ferromagnet, local-moment, spintronic, Kondo-lattice model

\maketitle

\section{Introduction}

Half-metals will play the key role in spintronics \cite{key-1}, as do 
semiconductors in current electronics. Differently to semiconductors however, 
the Fermi level in ideal half-metals lies in a gap between valence and conduction bands 
only for one spin direction, whereas for the other spin direction the bands
overlap. Thus the electrical current is exclusively due to spin polarized charge 
carriers, with promising consequences when exploiting the spin degree of freedom 
in logical devices. In semimetals, valence and conduction bands overlap regardless 
of the spin direction, leading to unpolarized currents.

The investigation of half-metallic ferromagnets started with density
functional theory calculations by de Groot et al. \cite{key-2} on
the Heusler alloy NiMnSb. Subsequently, other materials were identified
as half-metals, as e.g. other Heusler alloys such as $(Pt,Fe,Co)MnSb$
\cite{key-3}, ferromagnetic oxides as $CrO_{2}$\cite{key-4} and
$Fe_{3}O_{4}$ \cite{key-5}, and colossal magnetoresistance (CMR)
systems like $La_{1-x}Sr_{x}MnO_{3}$ \cite{key-6}.

There are no pure elements which are half-metallic ferromagnets. The
classical ferromagnets $Co$ and $Ni$ have fully polarized $3d$-states
at the Fermi edge, but there are also $4s$-states at $E_{F}$ preventing
a fully polarized current. A mechanism must be found either to push
the bottom of the $4s$-Band above the Fermi edge or to press the
latter below the band bottom. This is done by alloying or by forming
an oxidic compound. Thus all known half-metallic ferromagnets contain
more than one element.

An important issue is the explanation of the gap, the origin of which
is as equally diverse as the origin of half-metalicity. First-principles
bandstructure calculations \cite{key-2,key-7} predict such a gap
at $T=0$. What is the physical reason for it? What happens at finite
temperature? Magnon and phonon effects may give rise to a depolarization
so that strict half-metalicity appears to be limited to $T=0$?

In this paper we would like to model a local-moment half-metal to
understand the basic physics of the gap-behavior. Candidates for such
local-moment half-metals are the CMR-manganites, diluted ferromagnetic
semiconductors such as $Ga_{1-x}Mn_{x}As$ \cite{key-8,key-9}, and
even concentrated ferromagnetic semiconductors as $EuS$ \cite{key-9}.
We will describe these materials by a two-band Kondo-lattice model.
This model (also known as $sf$ model) describes the mutual influence
of two well-defined subsystems, localized magnetic moments stemming
from a partially filled electron shell (e.g. $4f$), and itinerant
electrons in a partially filled energy band. Both subsystems are coupled
by an on-site interband exchange interaction. One well-known consequence
of this mutual influence is the carrier-induced (anti)ferromagnetism
in the local-moment subsystem (RKKY; Rudermann-Kittel-Kasuya-Yosida),
and another is the striking temperature-dependence of the bandstates
resulting, e.g., in a red shift of the band bottom upon cooling below
$T_{C}$ \cite{key-10,key-11}.

To describe local-moment half-metals we apply a two-band Kondo-lattice
model (the principle idea is depicted in \fref{fig:model}). The 
itinerant electrons belong to a conduction and a valence
band. For simplicity both are considered s-bands. They are exchange
coupled to a system of localized magnetic moments. The exchange coupling is
ferromagnetic ($J^{c}>0$) for conduction electrons and antiferromagnetic
($J^{v}<0$) for valence electrons. We demonstrate how the temperature-dependence
of the bandstates can lead to a half-metallic groundstate with a transition
to semimetallic or semiconducting behavior with increasing temperature.
In a mean-field picture this is simply due to the fact that upon cooling,
the spin-up conduction band is energetically shifted downwards and
the spin-up valence band upwards, due to interband exchange between
the itinerant charge carriers and the localized moments. This results in 
an increased concentration of spin-up electrons. The opposite happens to the 
spin-down electrons. We investigate this important phenomena in terms of the 
local-moment magnetization and the electron spin polarization, and try to 
explain them via the special behavior of the quasiparticle-bandstructure.

In previous studies \cite{key-KLM1,key-KLM2,key-KLM3,key-KLM4,key-KLM5},
single-band Kondo-lattice models were used to describe half-metals. 
In this paper we investigate the two-band Kondo-lattice model, we had 
already proposed in \cite{key-eub6} to study the electronic properties of 
$EuB_6$. Now, we extend the analytical methods by the modified-RKKY 
approach \cite{key-11}, which allows us to calculate the 
magnetic properties self-consistently instead of considering the net 
magnetization as a parameter following the Brillouin function. 
Furthermore, the focus of this paper lies on the exploration of 
characteristics of the introduced model and the prediction of possible 
properties of local-moment half-metals. This is done in a general model 
study rather than a specific case study.

In the next section we introduce the model and outline the theory
we used for solving the underlying many-body problem. The third section
is then for a detailed discussion and interpretation of the results,
followed in the last chapter by some conclusions.

\section{Theory}

\begin{figure}

\begin{center}

\includegraphics[width=4in,clip]{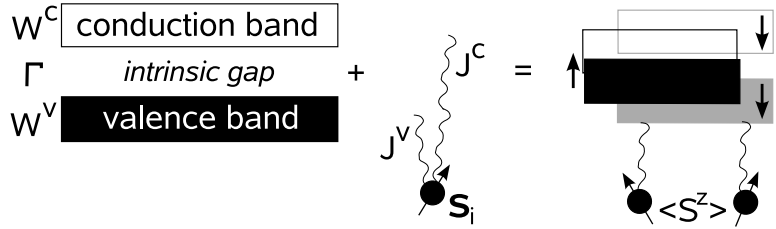}

\end{center}

\caption{\label{fig:model}From left to right: free system ($J=0$) with the
electronic parameters: bandwidths $W^{v}$ and $W^{c}$ of valence
and conduction bands, respectively and intrinsic band gap $\Gamma$; on-site $sf$ coupling $J^{v}<0$
and $J^{c}>0$ of the valence and conduction electrons with the local-moments
$\vec{S}_{i}$; half-metal with spin polarized electrons
at the Fermi edge and mean magnetization $\langle S^z \rangle$ of the local-moments}

\end{figure}

The proposed two-band Kondo-lattice model for half-metals describes
the electronic and magnetic systems which mutually influence each
other. The free electronic system (left picture in \fref{fig:model}) consists of two $s$-bands with widths $W^{v}$
and $W^{c}$ of valence and conduction bands, respectively. Between
these bands we consider an intrinsic band gap $\Gamma$ . As discussed
above, when switching on the magnetic interaction $J$ between the
spins of the electrons in the bands and the local magnetic moments
$\vec{S}_{i}$ (\fref{fig:model}, middle picture), the bands
will shift spin-dependently and form the half-metal (\fref{fig:model},
right picture). 

\subsection{Hamiltonian}

We describe the two-band Kondo-lattice model for local-moment
half-metals by the following Hamiltonian

\begin{equation}
H_{sf}=\sum\limits _{b}^{v,c}\left[\sum_{ij\,\sigma}\left(T_{ij}^{b}-\mu\delta_{ij}\right)a_{i\sigma}^{b\dagger}a_{j\sigma}^{b}-J^{b}\sum_{i}\vec{S}_{i}\cdot\vec{s}_{i}^{b}\right].\label{eq:twoband-Hamiltonian}
\end{equation}

\begin{itemize}

\item The index $b$ stands for valence ($v$) and conduction band ($c$),
$i$ and $j$ are different lattice sites and $\sigma=\uparrow$ or $\downarrow$ is the spin
index of the electrons. $a_{i\sigma}^{b}$ and $a_{i\sigma}^{b\dagger}$
are the annihilation and creation operators, respectively. The chemical
potential $\mu$ will be chosen to result in equal hole and electron
concentrations. 

\item The first term of the Hamiltonian \eref{eq:twoband-Hamiltonian}
is the kinetic part. $T_{ij}^{b}$ represent the hopping integrals
of the electrons and are connected to the Bloch energies $\varepsilon_{\vec{k}}$
in reciprocal space by Fourier transformation $T_{ij}^{b}=N^{-1}\sum_{\vec{k}}\varepsilon_{\vec{k}}^{b}\rme^{\rmi\vec{k}\cdot\vec{R}_{ij}}$.
For the Bloch energies we use the simple cubic, tight-binding energy
dispersion\cite{key-tb} 

\begin{equation}
\varepsilon_{\vec{k}}^{b} =z_{b}\left\{ \frac{\Gamma}{2}-\frac{W^{b}}{6}\left[\cos(k_{x}a)+\cos(k_{y}a)+\cos(k_{z}a)\right]\right\} ,\label{eq:tightbinding}
\end{equation}
where $k$ is the wavevector, $a$ the lattice constant and $z_{b}=\delta_{bc}-\delta_{bv}$
the band sign operator. The dispersion relations \eref{eq:tightbinding}
can be recognized in the rigidly shifted spin-down valence band and
spin-up conduction band in the top-most panels of \fref{fig:bandstructure}.

\item The second term of the Hamiltonian \eref{eq:twoband-Hamiltonian}
describes the on-site magnetic $sf$ coupling between the spins of the itinerant
band electrons $\vec{s}_{i}^{b}$ and the spin of the local-moments
$\vec{S}_{i}$. The coupling strength is $J=J^{c}=-J^{v}$, as already
discussed above. 
\end{itemize}

\subsection{Single electron Green functions}

For the solution of the many body problem the Green functions method
is used. From the Hamiltonian \eref{eq:twoband-Hamiltonian} we can
formally derive the single electron Green functions

\begin{equation}
G_{\vec{k}\sigma}^{b}(E)=\frac{\hbar}{E-\varepsilon_{\vec{k}}^{b}+\mu-\Sigma_{\sigma}^{b}(E)},\label{eq:greenfunction}
\end{equation}
where $b=v,c$ stands again for the valence and conduction bands,
respectively. The single electron Green functions \eref{eq:greenfunction}
are written down by use of the electron self-energy $\Sigma_{\sigma}^{b}(E)$.
Further details about the derivation of the self-energy for the Kondo-lattice
model can be found in \cite{key-isa,key-isa2}. The interpolating
self-energy approach \cite{key-isa} is sufficient for the
case of half metals, since we are only dealing with low
itinerant carrier densities.

For the low occupied conduction band, the self-energy from \cite{key-isa}
is used

\begin{equation}
\Sigma_{\sigma}^{c}(E)=M_{\sigma}^{c}+\left({\textstyle \frac{1}{2}}J^{c}\right)^{2}\frac{\alpha_{\sigma}G_{0}^{c}\left(E+M_{\sigma}^{c}\right)}{1-\frac{1}{2}J^{c}G_{0}^{c}\left(E+M_{\sigma}^{c}\right)},\label{eq:selfenergy_conductionband}
\end{equation}
with the abbreviation $\alpha_{\sigma}=S(1+S)-z_{\sigma}\langle S^{z}\rangle(1+z_{\sigma}\langle S^{z}\rangle)$,
the spin sign operator $z_{\sigma}=\delta_{\sigma\uparrow}-\delta_{\sigma\downarrow}$,
the mean magnetization $\langle S^{z}\rangle$, the free propagator
$G_{0}^{b}(E)=N^{-1}\sum_{\vec{k}}\hbar(E-\varepsilon_{\vec{k}}^{b}+\mu)^{-1}$
and what would be the mean field result for the self-energy $M_{\sigma}^{b}=-{\textstyle \frac{1}{2}}J^{b}z_{\sigma}\langle S^{z}\rangle.$

For the valence band we also have a low concentration of itinerant
carriers, the holes. The self-energy from \cite{key-isa} needs
some adaptation using the electron-hole symmetry. This was done previously in \cite{key-eub6}
and leads to the electron self-energy for the valence band

\begin{equation}
\Sigma_{\sigma}^{v}(E)=M_{\sigma}^{v}+\left({\textstyle \frac{1}{2}}J^{v}\right)^{2}\frac{\alpha_{-\sigma}G_{0}^{v}(E+M_{\sigma}^{v})}{1+\frac{1}{2}J^{v}G_{0}^{v}(E+M_{\sigma}^{v})}.\label{eq:selfenergy_valenceband}
\end{equation}

With the self-energies \eref{eq:selfenergy_conductionband} and \eref{eq:selfenergy_valenceband}
the Green functions \eref{eq:greenfunction} for valence and conduction
bands can now be determined. 

\subsection{Electronic properties}

The knowledge of the Green functions
allows the calculation of the spectral density

\begin{equation}
S_{\vec{k}\sigma}^{b}(E)=-{\textstyle \frac{1}{\pi}}\,{\rm Im}\, G_{\vec{k}\sigma}(E)\label{eq:spectraldensity}
\end{equation}
for any wavevector $\vec{k}$. Evaluated for wavevectors along the 
standard symmetry points of the first Brillouin zone \Eref{eq:spectraldensity} 
leads to the quasiparticle band structure presented in the next section.

Furthermore, we will present the spin polarization of the conduction
electrons 
\begin{equation}
P=\frac{n_{\uparrow}-n_{\downarrow}}{n_{\uparrow}+n_{\downarrow}},\label{eq:polarization}
\end{equation}
with the spin dependent electron density 
\begin{equation}
n_{\sigma}=\sum_{\vec{k}}\int f_{-}(E)S_{\vec{k}\sigma}^{c}(E)\, \rmd E\label{eq:occupationnumber}
\end{equation}
and the plasma frequency
\begin{equation}
\omega_{p}=\sqrt{\frac{e^{2}}{\epsilon_{0}}\frac{n_{\uparrow}+n_{\downarrow}}{m_{\uparrow}^{*}+m_{\downarrow}^{*}}}.\label{eq:plasma}
\end{equation}
The effective masses $m_{\sigma}^{*}$ are calculated via
\begin{equation}
\frac{m_{\sigma}^{*}}{m^{0}}=1-\left(\frac{\partial\,{\rm Re}\, G_{\vec{k}\sigma}^{c}(E_{\vec{k}\sigma}^{c})}{\partial E_{\vec{k}\sigma}^{c}}\right)_{\varepsilon_{\vec{k}}}\label{eq:effectivemass}
\end{equation}
at the $\Gamma$-point $\vec{k}=(000)$. In \Eref{eq:occupationnumber}
$f_{-}(E)=\left[{\rm exp}\left(\frac{E-\mu}{k_{\rm B}T}\right)+1\right]^{-1}$
denotes the Fermi function and in \Eref{eq:effectivemass} $E_{\vec{k}\sigma}^{c}=\varepsilon_{\vec{k}}-\mu+{\rm Re}\, G_{\vec{k}\sigma}^{c}(E_{\vec{k}\sigma}^{c})$
represent the quasiparticle resonance energies. The intrinsic effective
mass $m^{0}=\left(6\hbar^{2}\right)/\left(Wa^{2}\right)$ for $W=3.5eV$
and $a=5A$ equals half the free electron mass $m_{e}$.

\subsection{Magnetic properties}

For a consistent model study of the half-metals, we are also interested
in the magnetic properties. They are derived in the modified-RKKY
approach \cite{key-11}, where the Kondo lattice Hamiltonian
is mapped to an effective Heisenberg Hamiltonian and the 
magnetic properties are calculated self-consistently, using the
single electron Green functions. For the proposed two-band Kondo-lattice 
Hamiltonian \eref{eq:twoband-Hamiltonian} we will briefly recall the 
major steps in analogy to the derivation in \cite{key-11}.

The mapping of \Eref{eq:twoband-Hamiltonian} to an effective Heisenberg 
Hamiltonian is achieved by averaging out the itinerant $s$-electron degrees 
of freedom, but retaining the operator character of the local moment 
spin operators, yielding

\begin{equation}
H_{sf} \to \langle H_{sf} \rangle^{(s)} \equiv H_{ff}=-\sum_{ij}J_{ij}^{\rm{eff}}\vec{S}_{i}\cdot\vec{S}_{j}\label{eq:heisenberg_hamiltonian}
\end{equation}
In the averaging procedure the expectation value 
$\langle a_{\vec k + \vec q\, \sigma}^{b\dagger} \, a_{\vec k \, \sigma^{'}}^b \rangle^{(s)}$ 
is calculated in the electronic subspace. This is achieved by the restricted
Green function $G_{\vec k \sigma^{'} \, \vec k + \vec q \sigma}^{b(s)}$, 
for which the equation of motion can be written down exactly. The 
iteration of the equation of motion was possible but would lead to higher
orders of spin-operator products. Therefore, in the first iteration the single 
electron Green functions \eref{eq:greenfunction} are used for decoupling. This
leads to the effective Heisenberg exchange integrals as functionals of the
Kondo-lattice single electron Green functions

\begin{equation}
J_{ij}^{\rm{eff}}= \sum_{b}^{v,c}\left({\textstyle \frac{1}{2}}J^{b}\right)^{2}\int f_{-}(E)\,\frac{1}{\pi}\,{\rm Im} \left(G_{ij0}^{b}(E)\sum_{\sigma}G_{ij\sigma}^{b}(E)\right)\rmd E.\label{eq:heisenberg_exchangeintegrals}
\end{equation}
The Green functions \eref{eq:greenfunction} are here used in their
real space representation $G_{ij\sigma}^{b}(E)=N^{-1}\sum_{\vec{k}}G_{\vec{k}\sigma}^{b}(E)\rme^{\rmi\vec{k}\cdot\vec{R}_{ij}}$
and $G_{ij0}^{b}(E)$ stands for the Green function of the free system
($J=0$).

With the effective exchange integrals \eref{eq:heisenberg_exchangeintegrals}
the effective Heisenberg Hamiltonian \eref{eq:heisenberg_hamiltonian}
can now be used to calculate the magnetic properties in a standard procedure. 
We define a Green function of the local spin operators 
$\langle\langle S_i^+;\rme^{aS_j^z}S_j^- \rangle\rangle$, whose equation of motion
is decoupled using the Tyablikov-approximation. With the Callen method\cite{key-callen}
we eventually arrive at an equation for the mean magnetization
\begin{equation}
\langle S^{z}\rangle=\frac{(1+S+\varphi)\varphi^{2S+1}+(S-\varphi)(1+\varphi)^{2S+1}}{(1+\varphi)^{2S+1}-\varphi^{2S+1}},\label{eq:magnetization}
\end{equation}
where $\varphi=N^{-1}\sum_{\vec{q}}\left[{\rm exp}(\frac{E(\vec{q})}{k_{\rm B}T})-1\right]^{-1}$
is the mean magnon number and $E(\vec{q})=2\langle S^{z}\rangle\left(J^{\rm{eff}}_{\vec{q}=0}-J^{\rm{eff}}_{\vec{q}}\right)$
the magnon energies. The exchange integrals are here used in the reciprocal
space representation $J^{\rm{eff}}_{\vec{q}}=N^{-1}\sum_{ij}J^{\rm{eff}}_{ij}\rme^{-\rmi\vec{q}\cdot\vec{R}_{ij}}.$

Since the single electron Green functions \eref{eq:greenfunction}
are functionals of the mean magnetization \eref{eq:magnetization}
and vice versa, we get a closed system of equations which will be
solved self-consistently. The results from those calculations are
presented for one parameter set in the following section.

\section{Results}

The proposed two band Kondo lattice model \eref{eq:twoband-Hamiltonian}
implies a strong mutual influence between the electronic and magnetic
systems of the half-metals. On the one hand, the influence of the
local magnetic moments on the itinerant carriers - electrons in the
conduction band and holes in the valence band - leads to a magnetization
dependent band occupation. On the other hand, the magnetic exchange
coupling via the itinerant carriers (modified-RKKY; mRKKY) is a functional
of the band occupation itself, thus is the mean magnetization. This
mutual influence of the two subsystems is reflected in the different
phase transitions which will be presented in this section. 

For our qualitative description of ferromagnetic half-metals we will
reduce the parameter space through setting the band widths of the
valence and conduction bands to $W^{v}=W^{c}=3.5eV$, the local spin
quantum number to $S=\frac{7}{2}$ and the exchange coupling strengths
to $J^{c}=-J^{v}=0.5eV$. The variation of the intrinsic gap parameter
$\Gamma$ will shift the center of masses of valence and conduction
bands resulting in different regimes of band occupations, leading
to the different behavior of the half-metals. We will vary the intrinsic
gap exemplarily in the full range of possible ferromagnetism.

\subsection{Magnetization}

\begin{figure}[t]

\begin{center}
\includegraphics[width=3.5in,clip]{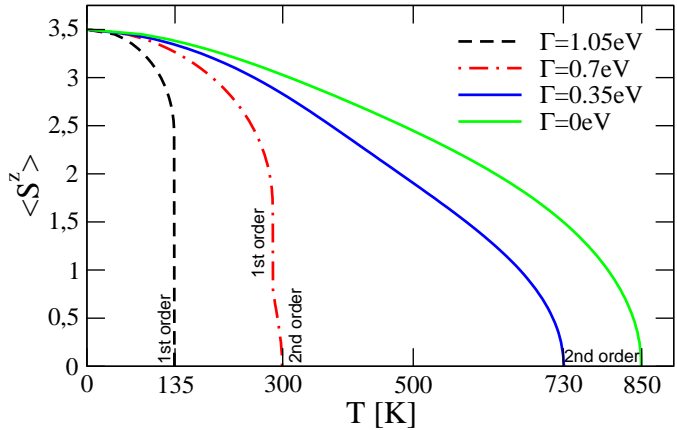}
\end{center}

\caption{\label{fig:magnetization}(color online) Temperature dependent magnetization curves
for different intrinsic band gaps $\Gamma$, fixed coupling strengths
$J^{c}=-J^{v}=0.5eV$, band widths $W^{c}=W^{v}=3.5eV$ and local
spin quantum number $S=\frac{7}{2}$. The solid lines show second
order phase transitions and the dashed line a first order phase transition
at $T_{C}$. The intermediate regime is represented by the dash-dotted
line which shows a first order phase transition close to $T_{C}$
and one of second order at $T_{C}.$ In the numeric calculations
\Eref{eq:magnetization} was solved self consistently, with the effective
Heisenberg exchange integrals \eref{eq:heisenberg_exchangeintegrals}
and the electron Green functions \eref{eq:greenfunction}.}
\end{figure}

The magnetization curves in \fref{fig:magnetization} are various
in their shapes due to the dynamic band occupation. Two different
regimes can be distinguished, namely those with magnetic phase transitions
of first order (dashed line) and second order (solid lines). Furthermore,
one parameter set (dash-dotted line) with a first as well as second
order phase transition is shown, representing the intermediate range.
The critical exponents of the magnetization at the second order 
phase transitions at the Curie temperatures $T_C$ are mean-field 
like ($\beta = 0.5\pm0.05$).

In all cases, a band overlap exists at low temperatures and the itinerant
carriers establish ferromagnetism. If, for increasing temperature,
the carrier concentration becomes too low to maintain ferromagnetism
via the RKKY-mechanism, the magnetization drops down suddenly (first
order phase transition - dashed line). If the carrier concentration
is sufficient up to the Curie temperature, the phase transition is
continuous (second order - solid lines). In the intermediate range
(dash-dotted line) the carrier concentration in the bands is not big
enough to hold up the RKKY-mechanism above a temperature of $285K$,
which leads to a sudden drop in the magnetization (first order transition).
Opposite to the case with even lower band occupation (dashed line),
the magnetization doesn't drop down to zero, since the band occupation
at temperatures between $285K$ and $300K$ is sufficient to hold
up a reduced magnetization. This leads to a the second order phase
transition at $T_{C}$. 

In the next section we will have a closer look at how exactly the band 
structure changes under the variation of the net magnetization.

\subsection{Bandstructure}

\begin{figure}[t]
\begin{center}
\includegraphics[width=4in,clip]{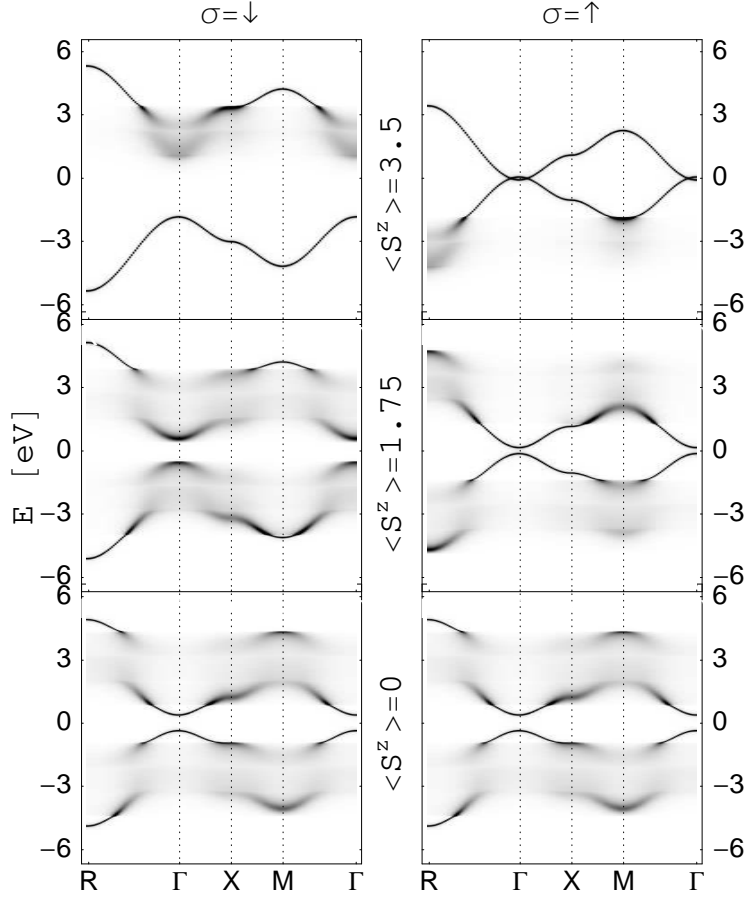}
\end{center}

\caption{\label{fig:bandstructure}Bandstructure for ferromagnetic saturation
at zero temperature (top), $\langle S^{z}\rangle=\frac{1}{2}S$ (middle)
and in the paramagnetic state (bottom) \cite{key-18}. The left-hand side shows the
spin-down bandstructure, the right-hand side spin-up. The parameter
are bandwidths $W^{c}=W^{v}=3.5eV$, intrinsic band gap $\Gamma=JS=1.75eV$,
sf-coupling strengths $J^{c}=-J^{v}=0.5eV$ and local spin quantum
number $S=\frac{7}{2}$. The Fermi level lies at $\mu=0eV$. The spectral 
density is plotted as grey scale for wavevectors along 
the standard symmetry points of the first Brillouin zone for simple cubic 
lattices (\Eref{eq:spectraldensity}).}
\end{figure}

The quasiparticle spectral density can be measured in (inverse) photoemission
experiments. 
In \fref{fig:bandstructure} we plot the spin dependent bandstructure as grey
scale of the spectral density \eref{eq:spectraldensity} at wavevectors along 
the standard symmetry points of the first Brillouin zone of simple cubic lattices.
The degree of blackening is a measure of the height of the spectral density, 
the width of the dispersion refers to the quasiparticle damping. A sharp
deep-black line refers to long-living quasiparticles. The representation
fits the bare line-shape of a respective (inverse) photoemission experiment.
The bandstructure is presented for three different mean magnetizations which 
were fixed during the calculations\cite{key-18}. The intrinsic gap $\Gamma=1.75eV$ in \fref{fig:bandstructure} is chosen for an easier discrimination
between valence and conduction bands. For the intrinsic gap parameters
used in \fref{fig:magnetization} the bandstructures would look qualitatively
the same as in \fref{fig:bandstructure}, only the bands would
be rigidly shifted towards each other by the difference in the intrinsic
band gap.

Before we come to a detailed discussion of the dynamic band occupation,
responsible for the different phase transitions, we will explain the
general properties that can be seen from the band structures of the
two-band Kondo-lattice model.

The upper right panel in \fref{fig:bandstructure} shows the
two spin up bands. The conduction band is the undeformed tight-binding
band. At zero temperature the crystal is ferromagnetically saturated
(all local-moments are aligned parallel). Thus no spin exchange processes
between the spin-up conduction electrons and the local-moments are
possible, and the band is only rigidly shifted by $-\frac{1}{2}JS$.
The same holds for the spin-down valence band (top left panel), due
to the more than half-filling of the valence band. 

The faded regions in the bandstructure result from scattering processes
of the electrons with the local-moments. Due to the scattering, the
lifetime of those states is finite. In the sharp regions the spectral
densities are delta functions and the quasiparticles have infinite
lifetime. They cannot take part in spin exchange processes with the
local-moments because no corresponding states with opposite spin are
available. Instead one can think of virtual scattering for those states.
At zero temperature the corresponding quasi particles are dressed
by virtual clouds of magnons called magnetic polarons. This solution
for the Kondo lattice model can be found analytically and is part
of the interpolating self energy approach\cite{key-isa} used for
\Eref{eq:selfenergy_conductionband} and \eref{eq:selfenergy_valenceband}.

With increasing temperature and decreasing magnetization (middle and
lower panels), the bands almost split into two flattened sub-bands.
The small overlap of spin-up valence and conduction bands decreases
with temperature and turns into a considerable bandgap in the paramagnetic
state (lower panels). 

As already mentioned above for different intrinsic band gaps, like
those in \fref{fig:magnetization}, the bands in \fref{fig:bandstructure}
would be rigidly shift towards each other resulting in greater band
overlaps. The band overlap still changes with magnetization but varies
in a different range. The different ranges are responsible for the
different magnetization curves seen in \fref{fig:magnetization}
since the effective direct exchange coupling between the local-moments
is mediated by the itinerant carriers (mRKKY-mechanism).

For intrinsic band gaps less than $0.7eV$, the band occupation is
sufficient to maintain the mRKKY-mechanism in the whole temperature
range. Thus the magnetization decreases continuously with increasing
temperature, and the magnetic phase transition at the Curie-temperature
is of second order (solid lines in \fref{fig:magnetization}).

For intrinsic band gaps greater than $0.7eV$ the band overlap, existent
for large mean magnetizations vanishes while the magnetization decreases.
Thus the band occupation goes to zero and the mRKKY-mechanism cannot
be upheld any longer. When reaching the critical temperature, the magnetization drops down suddenly
and we see a first order magnetic phase transition (dashed lines in \fref{fig:magnetization}).

For intrinsic band gaps around $0.7eV$, the magnetization curve in \fref{fig:magnetization} shows a first as well as second order
phase transition. In this case the band overlap is sufficient to hold
up a mean magnetization in the whole temperature range but it is not
continuous. At the first critical temperature the magnetization
drops down rapidly to continue at a smaller value continously up to
the Curie-temperature.

\subsection{Spin-polarization and electron density}

\begin{figure}[t]

\begin{center}
\includegraphics[width=3.5in,clip]{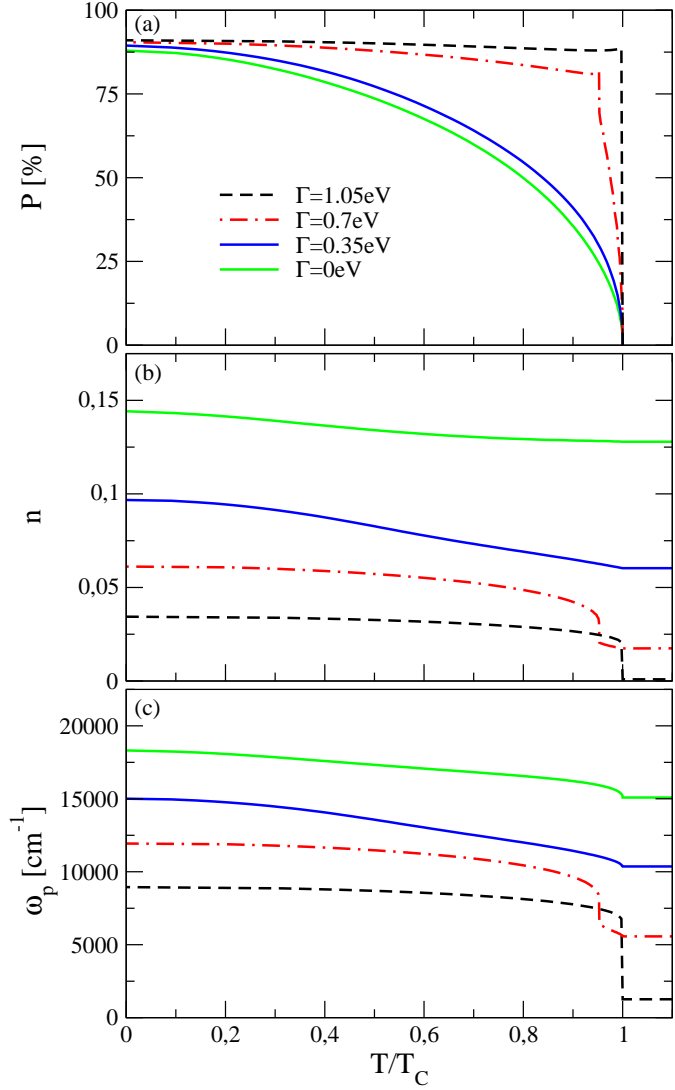}
\end{center}

\caption{\label{fig:polarization}(color online) (a)temperature dependent spin-polarization
$P$, (b) temperature dependent electron density $n$ and (c) temperature dependent plasma frequency $\omega_{p}$ for different intrinsic band gaps $\Gamma$, fixed coupling strengths
$J^{c}=-J^{v}=0.5eV$, band widths $W^{c}=W^{v}=3.5eV$ and local
spin quantum number $S=\frac{7}{2}$. The dashed line shows a transition
from half-metal to semiconductor and the other lines from half-metal
to semimetal.}
\end{figure}

The important property of half-metals is the spin polarization. 
In \fref{fig:polarization}a we present the spin polarization $P$, calculated from \Eref{eq:polarization} and in \fref{fig:polarization}b the electron density $n=n_{\uparrow}+n_{\downarrow}$, from \Eref{eq:occupationnumber}. The magnetization 
was calculated self-consistently as before.

One can see in the case $\Gamma=1.05eV$ (first order magnetic
phase transition), the spin polarization and electron density
drop down suddenly at $T_{C}$. Here, the sample undergoes a transition
from half-metal to semiconductor.

For the cases $\Gamma<0.7eV$ with second order magnetic phase transitions
we have a smoothly decreasing spin polarization. These samples undergo
a transition from half-metals to semimetals.

The case $\Gamma=0.7eV$, where a first and second order magnetic
phase transition is present, shows a corresponding drop of electron
density and spin-polarization at the same critical temperatures.

\subsection{Plasma frequency}

The plasma frequency can be obtained from optical absorption/reflectivity
measurements. 
\Fref{fig:polarization}c shows the plasma frequencies $\omega_p$, 
calculated from \Eref{eq:plasma} for the half-metals in discussion.
It depends on the electron density as well as the effective mass, which
both change with temperature in the introduced two-band Kondo-lattice
model. Although, the electron density varies more, compared
to the effective mass. Thus, the variation of the plasma 
frequency mainly depends on the temperature dependent electron density. Once 
more one can see the different behavior of the first and second order 
phase transitions as described above.

\section{Conclusion}

We introduced a two band Kondo lattice model which describes valence
and conduction electrons that are on-site coupled to local magnetic
moments. By propagating through the lattice, the electrons are responsible
for an effective direct exchange interaction (modified-RKKY) between
the local magnetic moments, which enables ferromagnetism.

The mutual influence between the spin dependent band overlap of valence
and conduction bands and the mean magnetization lead to different
phase transitions. On one hand it is possible to describe first as
well as second order magnetic phase transitions. On the other hand
one can explain the spin-electronic phase transitions: half-metal
to semimetal and half-metal to semiconductor.

For future spintronic device applications one could think of two possible
scenarios:

\begin{itemize}
\item We saw that changing the intrinsic bandgap would change the Curie-temperatures
and kinds of phase transitions. Applying e.g. hydrostatic pressure
on a sample, thus changing the lattice constant in a small range and
hence the intrinsic bandgap, could result in a variation of the electronic
and magnetic properties. This would be especially interesting around
the Curie-temperature. 
\item A stronger impact on the sample would have the application of an external
magnetic field at temperatures around the Curie-temperature. Depending
on the specific properties of the sample, one can think of magnetic
field-driven changes from the semiconducting (semimetallic) state
to a half-metallic state. 
\end{itemize}

In conclusion, first it is necessary to find appropriate materials
for which the theory is suitable. The search for that should focus
on compounds with the typical (ferromagnetic $J^c>0$, antiferromagnetic
$J^v<0$) Kondo lattice rare earths, such as $Ce, Eu$.
Although the theory was developed for periodic crystals with
local-moments on every lattice site, it is surely possible to expand
the model to randomly distributed local-moment compounds, which should
then show similar features. This possible future work would be interesting
for the diluted magnetic semiconductors (DMS).


\begin{thebibliography}{10}
\bibitem{key-18}The corresponding temperatures to the band structures 
in \fref{fig:bandstructure} can be read off from \fref{fig:magnetization}
(e.g. for $\Gamma=0.35eV$ the temperatures are $0K$, $530K$ and $730K$).
\end{thebibliography}
\end{document}